\documentclass[aps,prl,twocolumn,showpacs]{revtex4}

\usepackage{multirow}
\usepackage{graphicx}
\usepackage{dcolumn}
\usepackage{amsmath}
\usepackage{amssymb}
\usepackage{wasysym}
\usepackage{color}
\begin{document}

\title{Quantum oscillations from nodal bilayer magnetic breakdown in the underdoped high temperature superconductor YBa$_2$Cu$_3$O$_{6+x}$}

\author{Suchitra~E.~Sebastian$^1$}
\author{N.~Harrison$^2$}
\author{Ruixing Liang$^{3,4}$}
\author{D.~A.~Bonn$^{3,4}$}
\author{W.~N.~Hardy$^{3,4}$}
\author{C.~H.~Mielke$^2$}
\author{G.~G.~Lonzarich$^1$}

\affiliation{
$^1$Cavendish Laboratory, Cambridge University, JJ Thomson Avenue, Cambridge CB3~OHE, U.K\\
$^2$National High Magnetic Field Laboratory, LANL, Los Alamos, NM 87545\\
$^3$Department of Physics and Astronomy, University of British Columbia, Vancouver V6T 1Z4, Canada\\
$^4$Canadian Institute for Advanced Research, Toronto M5G 1Z8, Canada
}
\date{\today}

\begin{abstract}
We report quantum oscillations in underdoped YBa$_2$Cu$_3$O$_{6.56}$ over a significantly large range in magnetic field extending from $\approx$~24 to 101~T, enabling three well-spaced low frequencies at $\approx$~440~T, 532~T, and 620~T to be clearly resolved. We show that a small nodal bilayer coupling that splits a nodal pocket into bonding and antibonding orbits yields a sequence of frequencies, $F_0-\Delta F$, $F_0$ and $F_0+\Delta F$ and accompanying beat pattern similar to that observed experimentally, on invoking magnetic breakdown tunneling at the nodes. The relative amplitudes of the multiple frequencies observed experimentally in quantum oscillation measurements are shown to be reproduced using a value of nodal bilayer gap quantitatively consistent with that measured in photoemission experiments in the underdoped regime.
\end{abstract}
\maketitle

Discerning the electronic structure of underdoped YBa$_2$Cu$_3$O$_{6+x}$ in the normal state is a crucial step to understanding the origin of unconventional superconductivity in these materials~\cite{norman1}. While quantum oscillation measurements in underdoped YBa$_2$Cu$_3$O$_{6+x}$ have revealed multiple frequency components~\cite{sebastian3,audouard1,sebastian0}, it has been challenging to distinguish an electronic structure from the numerous possibilities that can uniquely explain the observed frequencies. Recent measurements of the chemical potential oscillations~\cite{sebastian1} in underdoped YBa$_2$Cu$_3$O$_{6+x}$ in strong magnetic fields have helped narrow down these possibilities by determining the multiple frequencies to arise from a single carrier pocket $-$ likely located at the nodal region of the Brillouin zone~\cite{sebastian4} and supported by other experiments sensitive to the density-of-states at the Fermi energy~\cite{riggs1,tranquada1,hossain1}. An intriguing question therefore arises as to how such a single pocket can give rise to the multiple observed frequencies.

\begin{figure} [htb!]
\centering 
\includegraphics*[width=.48\textwidth]{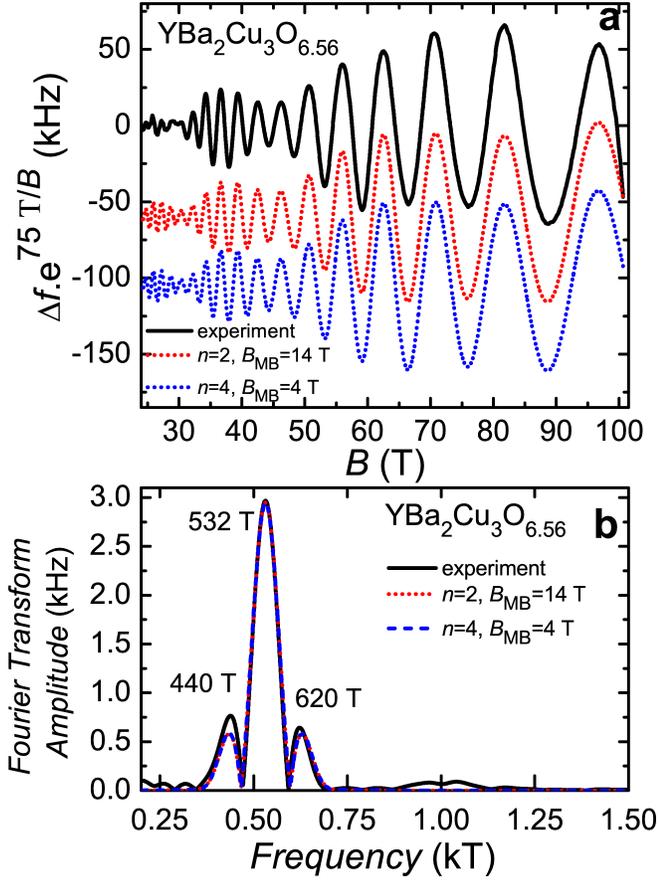}
\caption{{\bf a}, Magnetic quantum oscillations (solid line) measured in YBa$_2$Cu$_3$O$_{6.56}$ obtained using the contactless resistivity technique at $T=1.6(1)$~K consisting primarily of $\approx$~440~$\pm$~10~T, 532~$\pm$~2~T and 620~$\pm$~10~T frequencies (see {\bf b}). A slowly varying background has been subtracted, the data has been multiplied by the monotonic function $\exp[75 {\rm T}/B]$ for visual clarity. The upper (red) dotted curve is the simulated waveform for a nodal pocket split into bonding and antibonding orbits of frequency $F_0\pm\Delta F$ (using $F_0=$~530~T and $\Delta F=$~90~T from the experimental data) with $n=2$ magnetic breakdown gaps (see Fig.~\ref{nodalpocket}b) and a characteristic breakdown field $B_{\rm MB}\approx$~14~T. The lower (blue) dotted curve is the simulated waveform for a nodal pocket split into bonding and antibonding orbits of frequency $F_0\pm 2\Delta F$ with $n=4$ magnetic breakdown gaps (see Fig.~\ref{diamond}) and a characteristic breakdown field $B_{\rm MB}\approx$~4~T. Simulations are made using Equation~(\ref{qoamplitude}), in which the product $R_{\rm s}R_{\rm D}R_{\rm T}$~\cite{factors} is a monotonic function of $B$, and a negative prefactor is used for the $n=2$ case. {\bf b} Fourier transform (using a Blackman window) of the measured quantum oscillation data compared with simulated data. The three experimental low frequencies are seen to be reproduced by the simulation with the right ratio of amplitudes.}
\label{95T}
\end{figure}

In this paper, we show measurements of quantum oscillations in YBa$_2$Cu$_3$O$_{6+x}$ made using a contactless resistivity technique over an unprecedented range in magnetic field extending from $\approx$~24~T to 101~T (see Fig.~\ref{95T})~\cite{technique}. The large window in inverse magnetic field ($\Delta(\frac{1}{B})\approx$~0.032~T$^{-1}$) affords clear resolution of three well-separated low frequencies, namely $\approx$~440~$\pm$~10~T, 532~$\pm$~2~T and 620~$\pm$~10~T $-$ the experimental limit for distinguishing closely-spaced frequencies being $1/\Delta(\frac{1}{B})\approx$~32~T. We use the unique form of the multiple frequency quantum oscillation spectrum, in which the central frequency at 532~T is equidistantly flanked by two frequencies 532~$-$~90~T and 532~$+$~90~T, as a clue to infer a possible electronic structure that describes the system. We show that an electronic structure comprising a single bilayer-split nodal pocket combined with magnetic breakdown tunnelling would give rise to such a frequency spectrum, and potentially provide an explanation for the angular dependence of the observed frequencies~\cite{sebastian0,ramshaw1,sebastian2}. Similar instances of multiple frequencies arising from magnetic breakdown, are for instance found in heavy fermion and ferromagnetic families of materials~\cite{sigfusson1,taillefer1}. A prerequisite for such an explanation in underdoped YBa$_2$Cu$_3$O$_{6+x}$ is that a reduced bilayer splitting at the nodes~\cite{borisenko1} persists deep into the underdoped regime where the chains are partially occupied and a form of order~\cite{norman1,sebastian4} likely reconstructs the Fermi surface.

\begin{figure} [htb!]
\centering 
\includegraphics*[width=.46\textwidth]{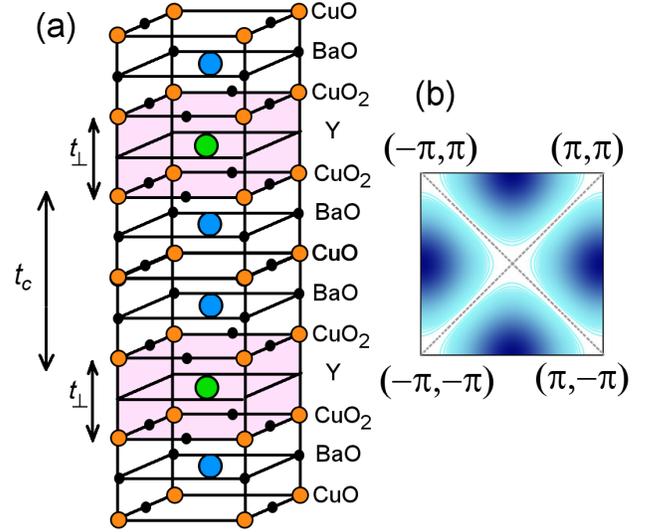}
\caption{{\bf} a, A schematic of two YBa$_2$Cu$_3$O$_{6+x}$ unit cells where the CuO$_2$ bilayers are shaded in pink. Hopping matrix elements within ($t_\perp$) and between ($t_c$) bilayers are shown. {\bf b}, A contour plot of $t_\perp({\bf k})$ within the CuO$_2$ planes for the simplified model expressed as equation~(\ref{inplanebilayer}), in which case the bilayer coupling vanishes at the nodes along which $|k_x|=|k_y|$ (shown by dotted gray lines) .}
\label{unitcell}
\end{figure}

The unit cell in YBa$_2$Cu$_3$O$_{6+x}$ contains a pair of CuO$_2$ layers, known as a bilayer (shown Fig.~\ref{unitcell}a). Coupling within a bilayer [denoted as $t_\perp({\bf k})$] splits bonding and antibonding bands by a finite gap~\cite{andersen1}. The general expectation, therefore, is for bilayer splitting to transform a single frequency $F_0$ of a single layer Fermi surface into two separate frequencies describing a bilayer Fermi surface, where the difference between them is proportional to the orbitally-averaged strength of the coupling $\bar{t}_\perp$ in the relevant region of the Brillouin zone. 

\begin{figure}
\centering 
\includegraphics*[width=.48\textwidth]{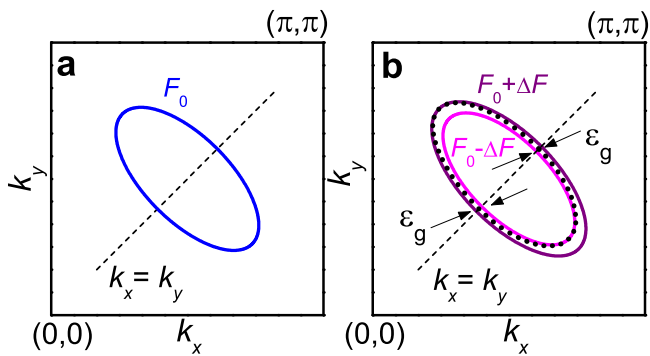}
\caption{A schematic nodal hole pocket before ({\bf a}) and after ({\bf b}) bilayer splitting, showing each of the allowed orbits with frequencies $F_0\pm\Delta F$, and the additional orbit allowed by magnetic breakdown tunnelling (dotted line) with frequency $F_0$. The location of the nodal magnetic breakdown gap $\varepsilon_{\rm g}$ is shown by arrows.}
\label{nodalpocket}
\end{figure}

The presence of three related low frequencies in the experimental data (Fig.~\ref{95T}) comprising two smaller-amplitude side frequencies equidistantly spaced from a central dominant-amplitude frequency seems difficult to reconcile with the simple expectation of two dominant-amplitude frequencies from bilayer splitting. An alternative explanation must be sought if the experimentally observed sequence of three frequencies is to be explained by a single carrier pocket. We find that such a possibility arises uniquely at the nodes in the underdoped regime, where recent doping-dependent photoemission measurements show the value of bilayer splitting ($\lessapprox$~16~meV) to be greatly reduced compared to that ($\approx$~150~meV) at the antinodes~\cite{fournier1}. The effect of this significant reduction in splitting at the nodes ($\varepsilon_{\rm g}$ in Fig.~\ref{nodalpocket}b) in the underdoped regime is to introduce the possibility of magnetic breakdown tunneling~\cite{shoenberg1}. We illustrate this in Fig.~\ref{nodalpocket} considering the example of a elliptical nodal hole pocket (such as would be created by an ordering vector of form ${\bf Q}=(\pi,\pi)$). Bilayer coupling splits the original pocket of frequency $F_0$ (Fig.~\ref{nodalpocket}a) into a bonding and antibonding orbit of frequency $F_0-\Delta F$ and $F_0+\Delta F$ respectively (Fig.~\ref{nodalpocket}b). The effect of the significant reduction in bilayer splitting at the nodes, is for a magnetic breakdown orbit (dotted line in Fig.~\ref{nodalpocket}b) to now become observable. As seen from Fig.~\ref{nodalpocket}a, this magnetic breakdown orbit has the same frequency ($F_0$) as the original unsplit Fermi surface. Tunneling across a reduced $\varepsilon_{\rm g}$ leads to the accumulation of spectral weight at $F_0$ at the expense of the adjacent frequencies $F_0-\Delta F$ and $F_0+\Delta F$. We thus arrive at a possible explanation for experimental observations in Fig.~\ref{95T} in the case where $F_0$ is associated with the central frequency at 532~T and the value of $\Delta F$ is taken to be 90~T, thereby yielding additional split frequencies at 440~T and 620~T.

The opening of a nodal gap ($\varepsilon_{\rm g}$ in energy) in YBa$_2$Cu$_3$O$_{6+x}$ found by angle-resolved photoemission studies~\cite{borisenko1} and supported by detailed electronic structure calculations~\cite{andersen1}, however, require us to go beyond simplified models of the coupling
\begin{eqnarray}
t_\perp({\bf k})=\frac{t_{\perp0}}{4}[\cos(k_xa)-\cos(k_yb)]^2
\label{inplanebilayer}
\end{eqnarray}
between bilayers which anticipate the splitting between bonding and antibonding bands to vanish at the nodes~\cite{andersen1,garcia1} (occurring along $|k_x|=|k_y|$ in Fig.~\ref{unitcell}b). Possible explanations for the opening of a nodal gap include a non vanishing nodal interbilayer coupling~\cite{garcia1, matrix} ($t_c$ in Fig.~\ref{unitcell}a), mixing with the chain bands~\cite{andersen1}, and certain forms of Fermi surface reconstruction~\cite{reconstructionnote}. 

To ascertain the anticipated relative amplitudes of $F_0-\Delta F$, $F_0$ and $F_0+\Delta F$ frequencies arising from nodal bilayer coupling and magnetic breakdown, we make use of the Falicov and Stachowiak treatment for magnetic breakdown amplitudes~\cite{shoenberg1}. An initial wave of unitary amplitude entering a magnetic breakdown junction is separated into a transmitted wave with amplitude $A_\nu=i\sqrt{P}$ and a reflected wave with amplitude $A_\eta=\sqrt{1-P}$, where the tunnelling probability ($P$) depends on the magnetic breakdown field $B_{\rm MB}$~\cite{breakdown} according to $P=e^{-B_{\rm MB}/B}$.

For multiple closed orbits, all possible closed orbits are additive, and the contribution from each orbit is multiplied by the magnetic breakdown reduction factor $R_{\rm MB}=(A_\nu)^{l_\nu}(A_\eta)^{l_\eta}$ where $l_\nu$ represents the number of magnetic breakdown points the orbit traverses by transmission, and $l_\eta$ represents the number of magnetic breakdown points the orbit traverses by reflection. The form of the total quantum oscillation amplitude, therefore, is yielded by 
\begin{eqnarray}
\sum_{m=-n/2}^{n/2}
N_{m}
R_{\rm MB}
R_{\rm s}
R_{\rm D}
R_{\rm T}
\cos{\bigg(\frac{2\pi (F_0 + m\Delta F)}{B}-\pi\bigg)}
\label{qoamplitude}
\end{eqnarray}
where $N_{m}$ enumerates closed orbits of the same frequency $F_0+m\Delta F$ (in the simple case of the elliptical nodal pocket considered, $n=l_\nu+l_\eta=2$, such that $m = -1,0,1$, and $N_0=2$, $N_{\pm 1}=1$ (table~\ref{table})), while $R_{\rm T}$, $R_{\rm D}$, and $R_{\rm s}$ represent the thermal, Dingle, and spin damping factors respectively~\cite{factors}.

We find that on introducing magnetic breakdown characterized by a single parameter $B_{\rm MB} \approx 14$~T to a simple bilayer-split nodal pocket (shown schematically in Fig.~\ref{nodalpocket}b), a waveform consisting of three frequencies $F_0-\Delta F$, $F_0$, and $F_0+\Delta F$ is produced that quantitatively explains two essential aspects of the experimental data: the field-dependence of the waveform in Fig.~\ref{95T}a, and the relative weights of the observed distribution of frequencies in Fig.~\ref{95T}b. The simulated data is produced using Equation~(\ref{qoamplitude}) in which the same value of $B_{\rm MB}$ generates different field-dependent probabilities for each of the three frequencies, tabulated in Table~\ref{table}. Such a value of magnetic breakdown field corresponds to a nodal gap ($\varepsilon_{\rm g}\approx\sqrt{(e\hbar B_{\rm MB}/m^\ast).\epsilon_{\rm F}}\approx$~10~meV~\cite{breakdown}), which falls within the upper limit ($\approx$~16~meV) measured for the nodal gap in photoemission experiments in the underdoped regime~\cite{fournier1}. This is in contrast to the significantly larger magnitude of antinodal gap of order $\varepsilon_{\rm g}~\approx~150$~meV experimentally measured by photoemission~\cite{fournier1}, which would preclude magnetic breakdown tunnelling, and yield only two split bilayer frequencies as discussed earlier. 
\begin{table}[h]
\begin{center}
\begin{tabular}{|l|l|l|l|l|} 
\hline
&\multicolumn{2}{|c|}{ellipse ($n=2$)}&\multicolumn{2}{|c|}{diamond ($n=4$)}\\
\hline
orbit&$N_m$&$R_{\rm MB}$&$N_m$&$R_{\rm MB}$\\
\hline
$F_0\pm2\Delta F$&$-$&$-$&1&$(1-P)^2$\\
\hline
$F_0\pm\Delta F$&1&$1-P$&4&$-P(1-P)$\\
\hline
$F_0$&2&$-P$&4&$-P(1-P)$\\
&&&2&$P^2$\\
\hline
\end{tabular}
\caption{\label{table} A table counting the number $N_m$ of instances each orbit $m$ of frequency $F_0+m\Delta F$ and cumulative magnetic breakdown probability $R_{\rm MB}$ occurs within the magnetic breakdown network. The bilayer-split elliptical (see Fig.~\ref{nodalpocket}b) and the diamond (see Fig.~\ref{diamond}) pocket scenarios are considered.}
\end{center}
\vspace{-0.6cm}
\end{table}

We next turn to the question of whether the reported warping of one of the observed Fermi surface pockets~\cite{audouard1,ramshaw1,sebastian0,sebastian2} is consistent with the bilayer splitting model proposed here. The splitting of the bonding and antibonding bands ($\varepsilon_{\rm a}$ and $\varepsilon_{\rm b}$) is given by~\cite{garcia1, matrix}
\begin{eqnarray}
\varepsilon({\bf k})_{{\rm a,b}}=\varepsilon({\bf k})\pm\sqrt{t_c^2+t_\perp^2+2t_ct_\perp\cos{(ck_z)}}
\label{bilayerperp}
\end{eqnarray}
(where $c$ is the $c$-axis lattice parameter, $k_z$ is the interlayer momentum and $\varepsilon({\bf k})$ is the single layer dispersion). For comparable values of $t_c$ and $t_\perp({\bf k})$ averaged over a nodal orbit, equation~\ref{bilayerperp} takes the form $\varepsilon({\bf k})_{{\rm a,b}}\approx\varepsilon({\bf k})\pm2t_\perp \cos{ck_z/2}$ . The modulation along $k_z$ of this dispersion results in an angular dependence of the bilayer-split frequencies $F_0\pm\Delta F$ which resembles that of a warped cylinder, potentially consistent with the experimentally observed angular dependence of the measured frequencies~\cite{sebastian0}. By contrast, the $k_z$ modulation is expected to be strongly suppressed in the case of a large disparity in magnitude between $t_\perp$ and $t_c$ as typically occurs at the antinodal region~\cite{andersen1}. We also note that the origin of the three frequencies ($F_0-\Delta F$, $F_0$ and $F_0+\Delta F$) from magnetic breakdown of a bilayer split pocket is consistent with their similar temperature-dependence reported in ref.~\cite{sebastian0}.

\begin{figure} [htp!]
\centering 
\includegraphics*[width=.46\textwidth]{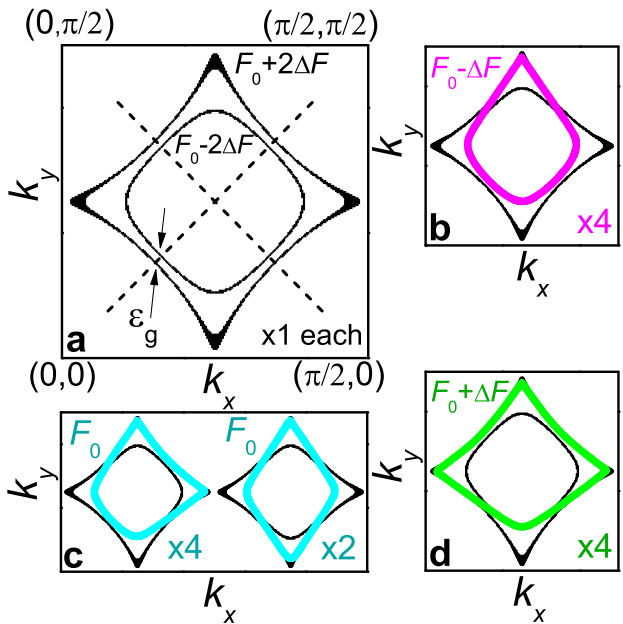}
\caption{A schematic of a diamond-shaped nodal pocket~\cite{harrison1,sebastian4} split by bilayer coupling, yielding 4 magnetic breakdown junctions with each of the possible orbits shown. (a) shows the two orbits (in this case $F_0-2\Delta F$ and $F_0+2\Delta F$) yielded by bilayer splitting of the original pocket, (b), (c), and (d) show additional allowed orbits due to nodal magnetic breakdown tunneling, yielding frequencies $F_0-2\Delta F$, $F_0 - \Delta F$, $F_0$, $F_0+\Delta F$ and $F_0+2\Delta F$ (the small breakdown field in the simulations of Fig.~\ref{95T} causes $F_0\pm2\Delta F$ to be weak in amplitude).}
\label{diamond}
\end{figure}

Finally, we note that the proposed scheme for multiple frequencies arising from a single carrier pocket by nodal bilayer coupling is generally valid for a pocket at the nodal region irrespective of its geometry or specific origin. We have thus far considered the simple case of elliptical pockets at the node (for instance, created by a spin-density wave or d-density wave~\cite{millis1,garcia1,sebastian3}) with two magnetic breakdown points. Another case which can be considered (e.g. Fig.~\ref{diamond}) is a nodal pocket with 4 magnetic breakdown points (in the shape of a diamond), as might result from bidirectional charge order~\cite{harrison1,sebastian4} or ortho-II band folding~\cite{podolsky1}. In this case, the multiplicity $N_m$ of orbits with frequencies $F_0-2\Delta F$, $F_0-\Delta F$, $F_0$, $F_0+\Delta F$, $F_0+ 2\Delta F$ is given by 1, 4, 6, 4 and 1 respectively (listed in table~\ref{table} as seen from orbit counting from Fig.~\ref{diamond}), corresponding to a modified binomial distribution. For this pocket geometry, the waveform and experimentally observed ratio of amplitudes of the  $F_0\pm \Delta F$ and $F_0$ frequencies is reproduced by a magnetic breakdown field of $B_{\rm MB}\approx 4$~T [simulation of quantum oscillation amplitude after Equation~(\ref{qoamplitude}) shown in Fig.~\ref{95T}]. The case of a nodal pocket of lower symmetry would yield multiple different values of $\Delta F$.

While previous interpretations of the measured three closely-spaced quantum oscillation frequencies attributed them to a nodal pocket accompanied by another pocket~\cite{sebastian0}, in the scenario presented here for bilayer-split frequencies, all three frequencies arise from a nodal pocket unaccompanied by another pocket; the latter rather than the former scenario is consistent with quantum oscillation experiments that point to a single carrier pocket~\cite{sebastian1}, in addition to the measured low value of measured heat capacity~\cite{riggs1}.

In summary, we report quantum oscillation data up to 101~T measured on YBa$_2$Cu$_3$O$_{6.56}$, providing a sufficiently wide range of inverse magnetic field to reveal a clear separation between three low frequencies at $\approx$~440~T, 532~T, and 620~T. Given the constraint that these multiple frequencies must arise from a single carrier pocket~\cite{sebastian1,sebastian4,riggs1}, here we propose an electronic structure involving bilayer splitting at the nodes that yields a hierarchy of frequencies similar to experimental observations. We show that for a value of magnetic breakdown field ($B_{\rm MB}\approx$~10~T) quantitatively consistent with the photoemission-measured magnitude of the nodal gap ($\varepsilon_{\rm g}\lessapprox$~16~meV~\cite{fournier1}) in the underdoped regime, a Fermi surface comprising a bilayer-split nodal pocket when acted on by magnetic breakdown tunneling explains two essential aspects of the data. It explains the field-dependent modulation of the waveform and the relative amplitudes of the observed sequence of frequencies $F_0 - \Delta F$, $F_0$ and $F_0+\Delta F$.


We acknowledge invaluable technical assistance from F.~Balakirev, J.~Betts, Y.~Coulter, M.~Gordon, R.~McDonald, D.~Rickel, and C.~Swenson. This work is supported by the US Department of Energy BES ``Science at 100 T," the National Science Foundation, the State of Florida, the Royal Society, and King's College (Cambridge University).


\begin{thebibliography}{99}
\bibitem{technique} In reference \cite{sebastian1}, quantum oscillations in the integrated contactless resistivity are shown to be proportional to those in the magnetic torque.
\bibitem{breakdown} The magnetic breakdown field is given by $B_{\rm MB}\approx\frac{m^\ast}{e\hbar}\frac{\epsilon_{\rm g}^2}{\epsilon_{\rm F}}$ where $\epsilon_{\rm F}=\hbar eF/m^\ast$ denotes the kinetic energy scale and $m^\ast$ is the quasiparticle effective mass.
\end{thebibliography}
\end{document}